\documentclass[12pt]{article}
\usepackage[cp1251]{inputenc}
\usepackage[T2A]{fontenc}
\usepackage[english]{babel}
\usepackage{amssymb}
\usepackage{amsmath}
\usepackage{amsfonts,mathrsfs,amscd}

\begin{document}
\title{On the spectrum of the Landau Hamiltonian perturbed by a periodic electric potential $V\in H^s_{\mathrm {loc}}({\mathbb R}^2;{\mathbb R})$, $s>0$}

\author{L.I.~Danilov}
\maketitle
\thispagestyle{empty}
\vspace{3mm}
\noindent{\it{\small Udmurt Federal Research Center of Ural Branch of the Russian Academy of Sciences, Izhevsk, Russia}}

\noindent{\it{\small danilov@udman.ru}}

\begin{abstract}
\noindent {\footnotesize  
We consider the Landau Hamiltonian $\widehat H_B+V$ on $L^2({\mathbb R}^2)$ perturbed by a periodic electric potential $V$. A homogeneous magnetic field $B>0$ is supposed to have the rational magnetic flux
$\eta =(2\pi )^{-1}Bv(K)\in {\mathbb Q}$ where $v(K)$ denotes the area of a unit cell $K$ of the period lattice $\Lambda $ of the potential $V$. We determine Banach spaces ${\mathcal L}^n_{\Lambda }({\mathbb R}^2;{\mathbb R})$ which are continuously embedded into Sobolev spaces $H^n_{\Lambda }({\mathbb R}^2;{\mathbb R})$, $n\in {\mathbb N} \cup \{ 0\} $, of $\Lambda $-periodic functions from $H^n_{\mathrm {loc}}({\mathbb R}^2;{\mathbb R})$, and contain dense $G_{\delta }$-sets ${\mathcal O}\subseteq {\mathcal L}^n_{\Lambda }({\mathbb R}^2;{\mathbb R})$ such that for every electric potential $V\in {\mathcal O}$ and every homogeneous magnetic field with the flux $0<\eta \in {\mathbb Q}$  the spectrum of the operator $\widehat H_B+V$ is absolutely continuous. In particular, the spaces ${\mathcal L}^n_{\Lambda }({\mathbb R}^2;{\mathbb R})=H^s_{\Lambda }({\mathbb R}^2;{\mathbb R})$, $s\in [n,n+1)$ can be chosen. For a given period lattice $\Lambda \subset {\mathbb R}^2$ and for a homogeneous magnetic field  
$B>0$ we also derive some conditions for Fourier coefficients of electric potentials $V\in H^n_{\Lambda }({\mathbb R}^2;{\mathbb R})$, $n\in {\mathbb N} \cup \{ 0\} $, under which in the case of the rational magnetic flux $\eta $ the spectrum of the operator $\widehat H_B+V$ is absolutely continuous.}\end{abstract}
\vskip 0.2cm
MSC 35P05

\section*{Introduction}

We consider the Landau Hamiltonian
$$
\widehat H_B+V=\left( -i\, \frac{\partial }{\partial x_1}\right) ^2+\left( -i\, \frac{\partial }{\partial x_2}-Bx_1\right) ^2+V
\eqno(1)
$$
acting on $L^2({\mathbb R}^2)$ where $B>0$ is the strength of a homogeneous magnetic field. The electric potential $V:{\mathbb R}^2\to {\mathbb R}$ is supposed to be periodic with a period lattice $\Lambda $. The coordanates in ${\mathbb R}^2$ are defined with respect to some orthonomal basis $e_1,e_2$. Let $E^1$ and $E^2$ be basis vectors of the lattice $\Lambda =\{ N_1E^1+N_2E^2: N_1,N_2\in {\mathbb Z}\} $. Let $K=\{ \xi _1E^1+\xi _2E^2:0\leqslant \xi _j\leqslant 1, j=1,2\} $ denote the unit cell of the lattice $\Lambda $ corresponding to the basis vectors $E^1,E^2$ and let $v(K)$ denote area of the unit cell $K$ (where $v(\cdot )$ is the Lebesgue measure on ${\mathbb R}^2$); $\eta =(2\pi )^{-1}Bv(K)$ is the (normalized) magnetic flux $K$ (which doesn't depend on the choice of the unit cell $K$). In the paper we assume that $\eta >0$ is a rational number ($\eta \in {\mathbb Q}$).

The operator (1) is a particular case of the two-dimensional Schr\"{o}dinger operator
$$
\left( -i\, \frac{\partial }{\partial x_1}-A_1\right) ^2+\left( -i\, \frac{\partial }{\partial x_2}-A_2\right) ^2+V
\eqno(2)
$$
where $A:{\mathbb R}^2\to {\mathbb R}^2$ and $V:{\mathbb R}^2\to {\mathbb R}$ are the magnetic potential and the electric potential. The magnetic field is defined by $B(x)=\frac{\partial A_2}{\partial x_1}-\frac{\partial A_1}{\partial x_2}\, $. The two-dimensional Schr\"{o}dinger operator (with periodic potentials $A$ and $V$ which have the common period lattice $\Lambda $) has been studied extensively (see [1--7] and the references given there). In [1] it was shown that the spectrum of the periodic operator (2) is absolutely continuous if $|A|\in L^p_{\mathrm {loc}}$, $p>2$, and $V\in L^q_{\mathrm {loc}}$, $q>1$. In the following papers the conditions on the potentials  $A$ and $V$ have been relaxed. In particular, in [6,7] it was proved that the spectrum of the periodic potential (2) is absolutely continuous if the functions $V$ and $|A|^2$ have bound zero in the sense of quadratic forms relative to the free Schr\"{o}dinger operator $-\Delta =-\frac {\partial ^2}{\partial x_1^2}-\frac {\partial ^2}{\partial x_2^2}\, $. The periodic Schr\"{o}dinger operators with a periodic variable metric and measure-derivative like electric potentials $V$ were also considered. For a survey of results on absolute continuity of the spectrum of multidimensional periodic Schr\"{o}dinger operators see [8--10].

Let $L^p_{\Lambda }({\mathbb R}^2;{\mathbb C})$, $p\in [1,+\infty )$, $L^{\infty }_{\Lambda }({\mathbb R}^2;{\mathbb C})$,
$C^n_{\Lambda }({\mathbb R}^2;{\mathbb C})$, $n\in {\mathbb Z}_+\doteq {\mathbb N} \cup \{ 0\} $, $C^{\infty }_{\Lambda }({\mathbb R}^2;{\mathbb C})$ and $H^s_{\Lambda }({\mathbb R}^2;{\mathbb C})$, $s\geqslant 0$, be the spaces of functions ${\mathcal W}:{\mathbb R}^2\to {\mathbb C}$ from $L^p_{\mathrm {loc}}({\mathbb R}^2)$, $L^{\infty }({\mathbb R}^2)$, $C^n({\mathbb R}^2)$, $C^{\infty }({\mathbb R}^2)$ and from Sobolev spaces $H^s_{\mathrm {loc}}({  \mathbb R}^2)$, respectively, which are periodic with period lattice $\Lambda $  ($C_{\Lambda }({\mathbb R}^2;{\mathbb C})=C^0_{\Lambda }({\mathbb R}^2;{\mathbb C})$ and $L^2_{\Lambda }({\mathbb R}^2;{\mathbb C})=H^0_{\Lambda }({\mathbb R}^2;{\mathbb C})$). The spaces of real-valued functions from these spaces are denoted analogously when replacing ${\mathbb C}$ with ${\mathbb R}$ in notation. For functions ${\mathcal W}\in L^p_{\Lambda }({\mathbb R}^2;{\mathbb C})$ and ${\mathcal W}\in C^n_{\Lambda }({\mathbb R}^2;{\mathbb C})$, $n\in {\mathbb Z}_+$, we define the norms $\| {\mathcal W}\| _{L^p_{\Lambda }}\doteq \| {\mathcal W}(\cdot |_K)\| _{L^p(K)}$ and
$$
\| {\mathcal W}\| _{C^n_{\Lambda }}\doteq \max\limits_{\mu ,{\,}\nu \, \in \, {\mathbb Z}_+ :\, \mu + \nu \, \leqslant \, n}\ \max\limits_{x\, \in \, K}\ \left| \frac{\partial ^{\, \mu +\nu }{\mathcal W}(x)}{\partial ^{\, \mu }x_1\, \partial ^{\, \nu }x_2}\right| \, .
$$

Let $|\cdot |$ and $(\cdot ,\cdot )$ denote the inner product and the length of vectors from ${\mathbb R}^2$. Let $E^1_*,E^2_*\in {\mathbb R}^2$ be vectors such that $(E^{\, \mu },E^{\, \nu }_*)=\delta _{\mu \nu }$, ${\mu },{\nu }=1,2$, where $\delta _{\mu \nu }$ is the Kronecker delta. The vectors $E^1_*$ and $E^2_*$ form a basis for the reciprocal lattice $\Lambda ^*=\{ N_1E^1_*+N_2E^2_*: N_1,N_2\in {\mathbb Z}\} $ 
with unit cell $K^*=\{ \xi _1E^1_*+\xi _2E^2_*: 0\leqslant \xi _j\leqslant 1, j=1,2\} $.

Let ${\mathcal W}_Y$ denote the Fourier coefficients of functions ${\mathcal W}\in L^2_{\Lambda }({\mathbb R}^2;{\mathbb C})$:
$$
{\mathcal W}_Y=(v(K))^{-1}\int_K{\mathcal W}(x)\, e^{-i(Y,x)}\, dx,\quad Y\in 2\pi \Lambda ^*.
$$

For ${\mathcal W}\in H^s_{\Lambda }({\mathbb R}^2;{\mathbb C})$ we define the norm
$$
\| {\mathcal W}\| _{H^s_{\Lambda }}\doteq (v(K))^{1/2}\, \Bigl( \, \sum\limits_{Y\, \in \, 2\pi \Lambda ^*}(1+|Y|)^{2s}|{\mathcal W}_Y|^2\, \Bigr) ^{1/2}.
$$

The spectrum of the Landau Hamiltonian
$$
\widehat H_B=\left( -i\, \frac{\partial }{\partial x_1}\right) ^2+\left( -i\, \frac{\partial }{\partial x_2}-Bx_1\right) ^2
$$
consists of the eigenvalues $\lambda =(2m+1)B$, $m\in {\mathbb Z}_+$, of infinite multiplicity (Landau levels). For a periodic point potential $V$, all Landau levels $\lambda =(2m+1)B$, $m\in {\mathbb Z}_+$, are eigenvalues of operator (1) if the period lattice $\Lambda $ is monatomic and $1<\eta \in {\mathbb Q}$ (and the spectrum also has an absolutely continuous component); see [11]. The electric potentials $V\in L^2_{\Lambda }({\mathbb R}^2;{\mathbb R})$ are $-\Delta $-bounded with relative bound zero (and therefore bounded with bound zero relative to the operators $\widehat H_B$) [12]. Hence, if $0<\eta \in {\mathbb Q}$, then the magnetic Floquet--Bloch theory applies (see [11]) and, for the electric potentials $V\in L^2_{\Lambda }({\mathbb R}^2;{\mathbb R})$, the spectrum of operator (1) has no singular component (see [11,13,14]). The absence of eigenvalues means the absolute continuity of the spectrum of the operator $\widehat H_B+V$.

For all nonconstant electric potentials 
$V\in L^2_{\Lambda }({\mathbb R}^2;{\mathbb R})$ with mean value $V_0=0$, if $\eta \in \{ Q^{-1}:Q\in {\mathbb N}\} $, then the spectrum of operator (1) has no eigenvalues outside the set $\{ (2m+1)B:m\in {\mathbb N}\} $ [15]. If $\eta \in \{ Q^{-1}:Q\in {\mathbb N}\} $ and either $V\in L^2_{\Lambda }({\mathbb R}^2;{\mathbb R})\backslash C^{\infty }_{\Lambda }({\mathbb R}^2;{\mathbb R})$ or $V:{\mathbb R}^2\to {\mathbb R}$ is a nonconstant trigonometric polynomial (with the period lattice $\Lambda $), then the spectrum of operator (1) is absolutely continuous [15,16].

For any $m\in {\mathbb N}$ and any homogeneous magnetic field $B>0$ there are nonconstant periodic electric potentials $V\in C^{\infty }({\mathbb R}^2;{\mathbb R})$ with mean value $V_0=0$ such that $\eta \in {\mathbb Q}$ and the Landau level $\lambda =(2m+1)B$ is an eigenvalue of the operator $\widehat H_B+V$ (see [17,18]).

It was shown in [19] that, for any period lattice $\Lambda \subset {\mathbb R}^2$, in the Banach space $(C_{\Lambda }({\mathbb R}^2;{\mathbb R}),\| \cdot \| _{C_{\Lambda }})$, there exits a dense $G_{\delta }$-set\footnote{The complement of the set ${\mathcal O}$ is a set of first Baire category.} ${\mathcal O}$ such that, for any potential $V\in {\mathcal O}$ and any homogeneous magnetic field with the flux $0<\eta \in {\mathbb Q}$, the spectrum of operator (1) is absolutely continuous. A similar result for potentials $V\in L^p_{\Lambda }({\mathbb R}^2;{\mathbb R})$, $p>1$, was obtained in [20]. The case $p=2$ was also considered in [21]\footnote{Theorem 3 in [21] asserts that there exists a set of second Baire category ${\mathcal O}\subseteq L^2_{\Lambda }({\mathbb R}^2;{\mathbb R})$ such that for all $V\in {\mathcal O}$ and all homogeneous magnetic fields with the flux $0<\eta \in {\mathbb Q}$ the spectrum of the operator $\widehat H_B+V$ is absolutely continuous. But in the proof the set ${\mathcal O}$ is obtained as the complement to a meager set, and so the set ${\mathcal O}$ containes a dense $G_{\delta }$-set.}. An analogous result for the relativistic Landau Hamiltonian (with a periodic electric potential) is given in [22].

The following two theorems are the main results in this paper.
\vskip 0.2cm

{\bf Theorem 1}. {\it Let $\Lambda $ be an arbitrary lattice in ${\mathbb R}^2$ and let $n\in {\mathbb Z}_+$. Suppose that, for a complex Banach spase $({\mathcal L}^n_{\Lambda }({\mathbb R}^2;{\mathbb C}),\| \cdot \| _{{\mathcal L}^n_{\Lambda }})$, the following four conditions are fulfilled: 

$\mathrm (a)$ \ the space ${\mathcal L}^n_{\Lambda }({\mathbb R}^2;{\mathbb C})$ is a linear subspace in the space $H^n_{\Lambda }({\mathbb R}^2;{\mathbb C})$,

$\mathrm (b)$ \ the space ${\mathcal L}^n_{\Lambda }({\mathbb R}^2;{\mathbb C})$ contains all trigonometric polynomials $\mathrm ($with the period lattice $\Lambda $$\mathrm )$,

$\mathrm (c)$ \ there exists a number ${\mathcal C}_{{\mathcal L}^n_{\Lambda }}>0$ such that, for all ${\mathcal W}\in {\mathcal L}^n_{\Lambda }({\mathbb R}^2;{\mathbb C})$,
$$
\| {\mathcal W}\| _{H^n_{\Lambda }}\, \leqslant \, {\mathcal C}_{{\mathcal L}^n_{\Lambda }}\, 
\| {\mathcal W}\| _{{\mathcal L}^n_{\Lambda }}\, ,
$$

$\mathrm (d)$ \
there exist numbers $\theta \in [0,1)$ and ${\mathcal C}^{\, \prime }_{{\mathcal L}^n_{\Lambda }}={\mathcal C}^{\, \prime }_{{\mathcal L}^n_{\Lambda }}(\theta )>0$ such that, for all $Y\in 2\pi \Lambda ^*$,
$$
\| e^{\, i(Y,x)}\| _{{\mathcal L}^n_{\Lambda }}\, \leqslant \, {\mathcal C}^{\, \prime }_{{\mathcal L}^n_{\Lambda }}\, (1+|Y|)^{n+\theta }.
$$
Then in the real Banach space $({\mathcal L}^n_{\Lambda }({\mathbb R}^2;{\mathbb R}),\| \cdot \| _{{\mathcal L}^n_{\Lambda }})$
$\mathrm ($which contains real-valued functions from the ${\mathcal L}^n_{\Lambda }({\mathbb R}^2;{\mathbb C})$$\mathrm )$ there is a dense $G_{\delta }$-set ${\mathcal O}$ such that, for every potential $V\in {\mathcal O}$ and every homogeneous magnetic field with the flux $0<\eta \in {\mathbb Q}$, the spectrum of the operator $\widehat H_B+V$ is absolutely continuous.}
\vskip 0.2cm

If $n=0$, the conditions (a)--(d) hold for the spaces $L^p_{\Lambda }({\mathbb R}^2;{\mathbb C})$, $p\geqslant 2$, and $
L^{\infty }_{\Lambda }({\mathbb R}^2;{\mathbb C})$. If $n\in {\mathbb Z}_+$, these conditions are fulfilled for the spaces $C^n_{\Lambda }({\mathbb R}^2;{\mathbb C})$ and $H^s_{\Lambda }({\mathbb R}^2;{\mathbb C})$, $s\in [n,n+1)$. Therefore, in particular, the following assertion is valid.
\vskip 0.2cm

{\bf Corollary 1}. 
{\it For each $s\geqslant 0$ there exists a dense
$G_{\delta }$-set ${\mathcal O}\subset (H^s_{\Lambda }({\mathbb R}^2;{\mathbb R}),\| \cdot \| _{H^s_{\Lambda }})$ such that for all potentials $V\in {\mathcal O}$ and all homogeneous magnetic fields with the flux $0<\eta \in {\mathbb Q}$ the spectrum of operator $\mathrm (1)$ is absolutely continuous.}
\vskip 0.2cm

For $B>0$, $V\in L^2_{\Lambda }({\mathbb R}^2;{\mathbb R})$ and $Y\in 2\pi \Lambda ^*$, we set
$$
{\mathcal C}_{B,V}(Y)\, \doteq \, |V_Y|\, -\, \sum\limits_{Y^{\prime }\, \in \, 2\pi \Lambda ^* \backslash \{ Y\} }|V_{Y^{\prime }}|\, e^{-|Y^{\prime }-Y|^2/(4B)}.
$$
\vskip 0.2cm

{\bf Theorem 2}.
{\it Let $V\in H^n_{\Lambda }({\mathbb R}^2;{\mathbb R})$, $n\in {\mathbb Z}_+$. Then the spectrum of the operator $\widehat H_B+V$ is absolutely continuous if $0<\eta \in {\mathbb Q}$ and 
$$
\limsup\limits_{2\pi \Lambda ^* \ni \, Y\to \, \infty }|Y|^{n+1}\, {\mathcal C}_{B,V}(Y)\, =\, +\infty .
$$}
\vskip 0.2cm

The particular case of Theorem 2 for $n=0$ was proved in [21].

In this paper we use the results and the methods from [15,16,21]. In \S{\,}1 we consider the magnetic Floquet--Bloch theory. The operator $\widehat H_B+V$ under the condition $\eta \in {\mathbb Q}$ is unitary equivalent to the direct integral of the ``fibrewise'' operators $\widehat H_B(k)+V$ depending on the magnetic quasimomentum $k$. Some assertions are given for operators $\widehat H_B(k+i\varkappa )$ and $\widehat H_B(k+i\varkappa )+V$ with $k+i\varkappa \in {\mathbb C}^2$. These assertions are used in \S{\,}2 to prove theorems 5 and 6 which are 'cornerstone' statements in the paper. Theorem 2 is a consequence of theorem 6. In \S{\,}3 we prove theorem 1.

\section{Magnetic Floquet--Bloch theory}

Operators (1) are unitary equivalent under different choice of basis vectors $e_1$ and $e_2$. Thus, the vectors $e_1,e_2$ and the basis vectors $E_1,E_2$ of the period lattice $\Lambda $ can be chosen so that $E^1_1>0$, $E^1_2=0$ and $E^2_2>0$, where $E^l_j=(E^l,e_j)$, $l,j=1,2$. Therefore $E^1_*=(E^1_1E^2_2)^{-1}(E^2_2e_1-E^2_1e_2)$, $E^2_*=(E^2_2)^{-1}e_2$ and $v(K^*)=(v(K))^{-1}=(E^1_1E^2_2)^{-1}$.

Let $\eta =PQ^{-1}$ where $P,Q\in {\mathbb N}$ are coprime integers. Instead of period lattice $\Lambda $ we consider the ``enlarged'' lattice $\widetilde {\Lambda }$ with basis vectors $\widetilde E^1 =QE^1$ and $\widetilde E^2=E^2$. The vectors $\widetilde E^1_*=Q^{-1}E^1_*$ and $\widetilde E^2_*=E^2_*$ are basis vectors of the lattice $\widetilde {\Lambda }^*$. We let $\widetilde K$ and $\widetilde K^*$ denote the unit cells of the lattices $\widetilde {\Lambda }$ and $\widetilde {\Lambda }^*$ corresponding to the basis vectors $\widetilde E^1,\widetilde E^2$ and $\widetilde E^1_*,\widetilde E^2_*$ respectively. For the unit cell $\widetilde K$, the magnetic flux satisfies the equality $(2\pi )^{-1}Bv(\widetilde K)=Q\eta =P\in {\mathbb N}$. Therefore, for operator (1) the magnetic Floquet--Bloch theory applies [11].

Let ${\mathcal H}^s_B$, $s\geqslant 0$, be the space of functions $\Phi :{\mathbb R}^2\to {\mathbb C}$ from $H^s_{\mathrm {loc}}({\mathbb R}^2;{\mathbb C})$ such that, for almost all $x\in {\mathbb R}^2$,
$$
\Phi (x+\widetilde E^{\mu })\, =\, e^{\, iB\widetilde E^{\mu }_1x_2}\, \Phi (x),\quad \mu =1,2. \eqno(3)
$$
We set ${\mathcal H}_B\doteq {\mathcal H}^0_B$ and ${\mathcal H}^{\infty }_B\doteq {\mathcal H}_B\cap C^{\infty }({\mathbb R}^2;{\mathbb C})$. Note that ${\mathcal H}^2_B\subset C({\mathbb R}^2;{\mathbb C})$. The space ${\mathcal H}_B$ is a Hilbert space equipped with the inner product
$$
(\Psi ,\Phi )_{{\mathcal H}_B}\, =\, \int_{\widetilde K}\overline {\Psi }\Phi \, dx,\quad \Psi ,\Phi \in {\mathcal H}_B.
$$
Let $\| \cdot \| _{{\mathcal H}_B}$ be the norm generated by this inner product, and let $\widehat I_B$ and $0_B$ be the identity operator on ${\mathcal H}_B$ and identically zero function from ${\mathcal H}_B$, respectively.

For functions $V\in L^2_{\Lambda }({\mathbb R}^2;{\mathbb R})$ (if $\eta =PQ^{-1}$) operator (1) is unitary equivalent to the direct integral
$$
\int _{2\pi \widetilde K^*}^{\oplus }(\widehat H_B(k)+V)\, \frac {dk}{(2\pi )^2v(\widetilde K^*)}\, 
\eqno(4) 
$$ 
acting on the space
$$
\int _{2\pi \widetilde K^*}^{\oplus }{\mathcal H}_B\, \frac {dk}{(2\pi )^2v(\widetilde K^*)}
$$ 
where
$$
\widehat H_B(k)=\left( k_1-i\, \frac{\partial }{\partial x_1}\right) ^2+\left( k_2-i\, \frac{\partial }{\partial x_2}-Bx_1\right) ^2, \quad k\in {\mathbb R}^2, 
$$
are self-adjoint operators on ${\mathcal H}_B$ with domain $D(\widehat H_B(k))={\mathcal H}^2_B$ [11,20]. In what follows we will also consider the operators $\widehat H_B(k+i\varkappa )$ for complex values of the magnetic quasimomentum $k+i\varkappa \in {\mathbb C}^2$, $k,\varkappa \in {\mathbb R}^2$. The potential $V\in L^2_{\Lambda }({\mathbb R}^2;{\mathbb R})$ (acting on ${\mathcal H}_B$ as an operator of multiplication) is $\widehat H_B(k)$-bounded with relative bound zero for all $k\in {\mathbb R}^2$. Hence the $\widehat H_B(k)+V$, $k\in {\mathbb R}^2$, are self-adjoint operators on ${\mathcal H}_B$ with domain $D(\widehat H_B(k)+V)=D(\widehat H_B(k))={\mathcal H}^2_B$. These operators have compact resolvents and discrete spectra.The decomposition into the direct integral (4) and Fredholm's analytic theorem implies the following theorem (see [11,13,16]).
\vskip 0.2cm

{\bf Theorem 3}.
{\it Let $V\in L^2_{\Lambda }({\mathbb R}^2;{\mathbb R})$ and $\eta =PQ^{-1}\in {\mathbb Q}$. Then $\lambda \in {\mathbb R}$ is an eigenvalue of operator $\mathrm (1)$ if and only if $\lambda $ is an eigenvalue of operators $\widehat H_B(k+i\varkappa )+V$ for all $k+i\varkappa \in {\mathbb C}^2$.}
\vskip 0.2cm

For each $k+i\varkappa \in {\mathbb C}^2$, the spectrum of the operator $\widehat H_B(k+i\varkappa )$ is discrete and coincides with the set of $P$-fold degenerate eigenvalues $\lambda =(2m+1)B$, $m\in {\mathbb Z}_+$ [11].

For vectors $k\in {\mathbb R}^2$, we denote by ${\mathcal H}^{(m)}_B(k)$, $m\in {\mathbb Z}_+$, the subspaces of eigenfunctions of the operator $\widehat H_B(k)$ with eigenvalues $\lambda =(2m+1)B$. Let define the operators
$$
\widehat Z_{\mp }(k)=\left( k_1-i\, \frac{\partial }{\partial x_1}\right) \pm i\left( k_2-i\, \frac{\partial }{\partial x_2}-Bx_1\right) 
$$
acting on ${\mathcal H}_B$ with domains 
$D(\widehat Z_{\mp }(k))={\mathcal H}^1_B$; $\widehat Z_{\mp }^*(k)=\widehat Z_{\pm }(k)$ and ${\mathcal H}^{(0)}_B(k)=\{ \Phi \in {\mathcal H}^1_B:\widehat Z_-(k)\Phi =0_B\} $. The following equalities hold:
$$
\widehat H_B(k)=\widehat Z_+(k)\widehat Z_-(k)+B=\widehat Z_-(k)
\widehat Z_+(k)-B .
$$
Given functions $\psi (k)\in {\mathcal H}^{(0)}_B(k)$, we set
$$
\psi ^{(m)}(k)=\frac {(2B)^{-m/2}}{\sqrt {m!}}\, \widehat Z^m_+(k)\psi (k)\in {\mathcal H}^{(m)}_B(k),\quad m\in {\mathbb Z}_+ .
$$
Note that
$$
\widehat Z_+(k)\psi ^{(m)}(k)=\sqrt {2B(m+1)}\, \psi ^{(m+1)}(k), \quad m\in {\mathbb Z}_+ ,
$$ $$
\widehat Z_-(k)\psi ^{(m)}(k)=\sqrt {2Bm}\, \psi ^{(m-1)}(k), \quad m\in {\mathbb N}. 
$$
If $\psi _j(k)$, $j=1,\dots ,P$, is orthonomal basis of the subspace ${\mathcal H}^{(0)}_B(k)$, then $\psi ^{(m)}_j(k)$, $j=1,\dots ,P$, is orthonomal basis of the subspace ${\mathcal H}^{(m)}_B(k)$, $m\in {\mathbb Z}_+$, and $\psi ^{(m)}_j(k)$, $j=1,\dots ,P$, 
$m\in {\mathbb Z}_+$, is orthonomal basis of the space ${\mathcal H}_B$.

The following lemma is a particular case of Lemma 3.5 in [19]. 
\vskip 0.2cm

{\bf Lemma 1}.
{\it For all $k\in {\mathbb R}^2$ and $\psi (k)\in {\mathcal H}^{(0)}_B(k)$,
$$
\| \psi (k)\| _{L^{\infty }({\mathbb R}^2)}\, \leqslant \, C_1 \, \| \psi (k)\| _{{\mathcal H}_B}
$$
where $C_1=C_1(\Lambda ,B,\widetilde K)>0$.}
\vskip 0.2cm

Let $\widehat P^{(m)}(k)$, $m\in {\mathbb Z}_+$, be an orthogonal projection of ${\mathcal H}_B$ onto the subspace ${\mathcal H}^{(m)}_B(k)$. 
For all $k\in {\mathbb R}^2$, we have
$$
{\mathcal H}^n_B=\left\{ \Phi \in {\mathcal H}_B:\sum\limits_{m\, =\, 1}^{+\infty }m^n\, \| \widehat P^{(m)}(k)\Phi \| ^2_{{\mathcal H}_B}<+\infty \right\} ,\quad n\in {\mathbb Z}_+.
$$
\vskip 0.2cm

{\bf Lemma 2}.
{\it For any $k\in {\mathbb R}^2$ and any $Y\in 2\pi \Lambda ^*$, there exists a unitary operator $\widehat U^{(Y)}(k)$ acting on the subspace ${\mathcal H}^{(0)}_B(k)$, such that, for all $\psi ,\psi ^{\prime }\in {\mathcal H}^{(0)}_B(k),$
$$
(\psi ^{\prime },e^{\, i(Y,x)}\psi )_{{\mathcal H}_B}\, =\, e^{-|Y|^2/(4B)}\, (\psi ^{\prime },\widehat U^{(Y)}(k)\psi )_{{\mathcal H}_B}.
$$}
\vskip 0.2cm

Lemma 2 was proved in [16]. 

We have
$$
\widehat H_B(k+(\zeta /2)e_1\pm i(\zeta /2)e_2)\, =\, \widehat Z_+(k)\widehat Z_-(k)+\zeta \widehat Z_{\mp }(k)+B, \quad \zeta \in {\mathbb C}. \eqno(5)
$$
\vskip 0.2cm

{\bf Lemma 3}.
{\it For all $k\in {\mathbb R}^2$ and all $\Phi \in {\mathcal H}^2_B$ such that $\widehat P^{(0)}(k)\Phi =0_B$, the estimate 
$$
\| (\widehat Z_+(k)\widehat Z_-(k)+\zeta \widehat Z_-(k))\Phi \| _{{\mathcal H}_B}\, \geqslant \, \sqrt {B/2}\ |\zeta |\, \| \Phi \| _{{\mathcal H}_B}
\eqno(6) 
$$
holds.}
\vskip 0.2cm
 
The proof of Lemma 3 is given in [21]\footnote{Estimate (6) was also proved in [16] for some constant $C>0$ in place of $\sqrt {B/2}$}. The following theorem was proved in [21] for some fixed vector $k\in 2\pi \widetilde K^*$ (which can be chosen in what follows) and was extended to all vectors $k\in {\mathbb R}^2$ in [15].  
\vskip 0.2cm

{\bf Theorem 4}.
{\it Let ${\mathcal W}\in L^2_{\Lambda }({\mathbb R}^2;{\mathbb R})$ and suppose that the magnetic flux for a homogeneous magnetic field satisfies the condition $\eta =PQ^{-1}\in {\mathbb Q}$. Then for any $\varepsilon >0$ there is a number $T({\mathcal W},\varepsilon )=T(\Lambda ,B,\widetilde K;{\mathcal W},\varepsilon )>0$ such that, for all $\zeta \in {\mathbb C}$ with $|\zeta |\geqslant T({\mathcal W},\varepsilon )$, all $k\in {\mathbb R}^2$, and all $\Phi \in {\mathcal H}^2_B$ for which $\widehat P^{(0)}(k)\Phi =0_B$, the inequality 
$$
\| {\mathcal W}\Phi \| _{{\mathcal H}_B}\, \leqslant \varepsilon \, \| (\widehat Z_+(k)\widehat Z_-(k)+\zeta \widehat Z_-(k))\Phi \| _{{\mathcal H}_B}.
$$
is fulfilled.}
\vskip 0.2cm
 
\section{Proof of Theorem 2}

For a given lattice $\Lambda \subset {\mathbb R}^2$ and a homogeneous magnetic field with the flux $\eta =PQ^{-1}$, we fix a vector $k\in 2\pi \widetilde K^*$ (the chosen unit cells $\widetilde K$ and $\widetilde K^*$ of the lattices $\widetilde {\Lambda }$ and $\widetilde {\Lambda }^*$ correspond to basic vectors $\widetilde E^1,\widetilde E^2$ and $\widetilde E^1_*,\widetilde E^2_*$, respectively). On the space ${\mathcal H}^n_B$, $n\in {\mathbb Z}_+$, let us define the metric
$$
\| \Phi \| _{k,{\mathcal H}^n_B}\, \doteq \, \| \widehat Z_+^n(k)\Phi \| _{{\mathcal H}^n_B}.
$$
If $\Phi \in {\mathcal H}^n_B$, $n\in {\mathbb N}$, then $\widehat Z_{\mp }(k)\Phi \in {\mathcal H}^{n-1}_B$ and
$$
\| \widehat Z_-(k)\Phi \| _{k,{\mathcal H}^{n-1}_B}\, \leqslant \, \| \widehat Z_+(k)\Phi \| _{k,{\mathcal H}^{n-1}_B}\, =\, \| \Phi \| _{k,{\mathcal H}^n_B}.
\eqno(7)
$$
Let $\widehat Z_-^{-1}(k)$ be the right inverse of the operator $\widehat Z_-(k)$ such that, for all $\psi (k)\in {\mathcal H}^{(0)}_B(k)$ and $m\in {\mathbb Z}_+$,
$$
\widehat Z_-^{-1}(k)\psi ^{(m)}(k)\, =\, (2B(m+1))^{-1/2}\, \psi ^{(m+1)}(k).
$$
If $\Phi \in {\mathcal H}^n_B$, $n\in {\mathbb Z}_+$, then $\widehat Z_-^{-1}(k)\Phi \in {\mathcal H}^{n+1}_B$ and
$$
\| \widehat Z_-^{-1}(k)\Phi \| _{k,{\mathcal H}^{n+1}_B}\, = \, \| \widehat Z_+^{n+1}(k)\widehat Z_-^{-1}(k)\Phi \| _{{\mathcal H}_B}\, \leqslant 
\eqno(8)
$$ $$
\leqslant \, \sqrt {(n+1)(n+2)}\, \| \widehat Z_+^n(k)\Phi \| _{{\mathcal H}_B}\, =\, 
\sqrt {(n+1)(n+2)}\, \| \Phi \| _{k,{\mathcal H}^n_B}.
$$

For any lattice $\Lambda ^{\prime }\subset {\mathbb R}^2$ the embeddings $H^1_{\Lambda ^{\prime }}({\mathbb R}^2;{\mathbb C})\subset L^4_{\Lambda ^{\prime }}({\mathbb R}^2;{\mathbb C})$ and
$H^{n+2}_{\Lambda ^{\prime }}({\mathbb R}^2;{\mathbb C})\subset C^n_{\Lambda ^{\prime }}({\mathbb R}^2;{\mathbb C})$, $n\in {\mathbb Z}_+$, are continuous. Hence the following mappings are also continuous:
$$
H^1_{\Lambda ^{\prime }}({\mathbb R}^2;{\mathbb C})\times H^2_{\Lambda ^{\prime }}({\mathbb R}^2;{\mathbb C})\ni ({\mathcal F},{\mathcal G})\mapsto {\mathcal {FG}}\in H^1_{\Lambda ^{\prime }}({\mathbb R}^2;{\mathbb C}),
$$ $$
H^n_{\Lambda ^{\prime }}({\mathbb R}^2;{\mathbb C})\times H^n_{\Lambda ^{\prime }}({\mathbb R}^2;{\mathbb C})\ni ({\mathcal F},{\mathcal G})\mapsto {\mathcal {FG}}\in H^n_{\Lambda ^{\prime }}({\mathbb R}^2;{\mathbb C}),\quad n\geqslant 2.
$$
Since a lattice $\Lambda ^{\prime }$ is arbitrary, it follows from these embeddings that Lemmas 4 and 5 are true (see, for example, the proof of estimate (2.13) in [22]).
\vskip 0.2cm

{\bf Lemma 4}.
{\it For all ${\mathcal W}\in H^1_{\Lambda }({\mathbb R}^2;{\mathbb C})$ and all $\Phi \in {\mathcal H}^2_B$ the inclusion ${\mathcal W}\Phi \in {\mathcal H}^1_B$ holds and
$$
\| {\mathcal W}\Phi \| _{k,{\mathcal H}^1_B}\, \leqslant \, C_3\, \| {\mathcal W}\| _{H^1_{\Lambda }}\, \| \Phi \| _{k,{\mathcal H}^2_B} ;
$$
if ${\mathcal W}\in H^2_{\Lambda }({\mathbb R}^2;{\mathbb C})$ and $\Phi \in {\mathcal H}^1_B$, then we also have ${\mathcal W}\Phi \in {\mathcal H}^1_B$ and
$$
\| {\mathcal W}\Phi \| _{k,{\mathcal H}^1_B}\, \leqslant \, C_3\, \| {\mathcal W}\| _{H^2_{\Lambda }}\, \| \Phi \| _{k,{\mathcal H}^1_B} 
$$
where $C_3=C_3(\Lambda ,B,\widetilde K^*)>0$.}
\vskip 0.2cm

{\bf Lemma 5}.
{\it For all ${\mathcal W}\in H^n_{\Lambda }({\mathbb R}^2;{\mathbb C})$ and all $\Phi \in {\mathcal H}^n_B$, $n\geqslant 2$ the inclusion ${\mathcal W}\Phi \in {\mathcal H}^n_B$ holds and
$$
\| {\mathcal W}\Phi \| _{k,{\mathcal H}^n_B}\, \leqslant \, C_4(n)\, \| {\mathcal W}\| _{H^n_{\Lambda }}\, \| \Phi \| _{k,{\mathcal H}^n_B} 
$$
where $C_4(n)=C_4(\Lambda ,B,\widetilde K^*;n)>0$.}
\vskip 0.2cm

Suppose that the operator $\widehat H_B+V$ for a potential $V\in H^n_{\Lambda }({\mathbb R}^2;{\mathbb R})$, $n\in {\mathbb Z}_+$, has an eigenvalue $\lambda \in {\mathbb R}$. Then it follows from (5) and Theorem 3 that the operators $\widehat Z_+(k)\widehat Z_-(k)+\zeta \widehat Z_-(k)+V$, for all $\zeta \in {\mathbb C}$, have the eigenvalue $\lambda -B$. Denote by $\Phi _{\zeta }\in {\mathcal H}^2_B$, $\zeta \in {\mathbb C}$, some eigenfunctions of these operators with the eigenvalues $\lambda -B$. Let $\Psi _{\zeta }=\widehat P^{(0)}(k)\Phi _{\zeta }$. Lemmas 1, 3 and Theorem 4 imply that, for $|\zeta |\geqslant T(V+B-\lambda;1/2)$,
$$
C_1\sqrt Q\, \| V+B-\lambda \| _{L^2_{\Lambda }}\, \| \Psi _{\zeta }\| _{{\mathcal H}_B}\,
=\, C_1\, \| V(\cdot |_K)+B-\lambda \| _{L^2(\widetilde K)}\, \| \Psi _{\zeta }\| _{{\mathcal H}_B}\, \geqslant
$$ $$
\geqslant \, \| (V+B-\lambda )\Psi _{\zeta }\| _{{\mathcal H}_B} \, =\| (\widehat Z_+(k)\widehat Z_-(k)+\zeta \widehat Z_-(k)+V+B-\lambda )(\widehat I_B-\widehat P^{(0)}(k))\Phi _{\zeta }\| _{{\mathcal H}_B}\, \geqslant
$$ $$
\geqslant \frac 12\, \| (\widehat Z_+(k)\widehat Z_-(k)+\zeta \widehat Z_-(k))(\widehat I_B-\widehat P^{(0)}(k))\Phi _{\zeta }\| _{{\mathcal H}_B}\, \geqslant
\, \frac {\sqrt B}{2\sqrt 2}\, |\zeta |\, \| (\widehat I_B-\widehat P^{(0)}(k))\Phi _{\zeta }\| _{{\mathcal H}_B}.
$$
Hence $\| \Psi _{\zeta }\| _{{\mathcal H}_B}\neq 0$ and
$$
\| (\widehat I_B-\widehat P^{(0)}(k))\Phi _{\zeta }\| _{{\mathcal H}_B}\, \leqslant \, 2\sqrt 2\, C_1Q^{\, 1/2}B^{-1/2}\, \| V+B-\lambda \| _{L^2_{\Lambda }}\, |\zeta |^{-1}\, \| \Psi _{\zeta }\| _{{\mathcal H}_B}.
$$
In what follows we assume that $\| \Psi _{\zeta }\| _{{\mathcal H}_B}=1$, $|\zeta |\geqslant T(V+B-\lambda ;1/2)$.
\vskip 0.2cm

{\bf Theorem 5}.
{\it There exists a constant $C_5(n)=C_5(\Lambda ,B,\widetilde K^*;n)>0$ such that, for all $\zeta \in {\mathbb C}$ with $\zeta =-Y_1+iY_2$, where $Y\in 2\pi \Lambda ^*$, and $|\zeta |\geqslant T(V+B-\lambda ;1/2)$, the estimate
$$
|(e^{\, i(Y,x)}\, \Psi _{\zeta },(V+B-\lambda )\Psi _{\zeta })_{{\mathcal H}_B}|\, \leqslant \, C_5(n)\bigl( 1+\| V+B-\lambda \| _{H^n_{\Lambda }}\bigr) ^{n+2}\, |\zeta |^{-n-1}.
\eqno(9)
$$
holds.}
\vskip 0.2cm

{\it Proof}.
For $\zeta \in {\mathbb C}$ with $|\zeta |\geqslant T(V+B-\lambda ;1/2)$, let's represent the eigenfunction $\Phi _{\zeta }$ as
$$
\Phi _{\zeta }\, =\, \Psi _{\zeta }+\zeta ^{-1}\, \Phi _{(1)}+ \dots +\zeta ^{-n}\, \Phi _{(n)}+\zeta ^{-n-1}\, \widetilde {\Phi }_{(n+1)}
$$
where the functions $\Phi _{(j)}$, $j=1,\dots ,n$, are sequentially determined from conditions $\widehat P^{(0)}(k)\Phi _{(j)}=0_B$ and (assuming $\Phi _{(0)}\doteq \Psi _{\zeta }$)
$$
(\widehat Z_+(k)\widehat Z_-(k)+V+B-\lambda )\, \Phi _{(j-1)}\, =\, -\widehat Z_-(k)\, \Phi _{(j)}, \quad j=1,\dots ,n.
$$
If $n=0$, then $\Phi _{\zeta }\, =\, \Psi _{\zeta }+\zeta ^{-1}\, \widetilde {\Phi }_{(1)}$. From (7), (8) and Lemmas 4 and 5, it follows that
$$
\Phi _{(j)}=(-1)^j\widehat Z_-^{-1}(k)\bigl( \widehat Z_+(k)+(V+B-\lambda )\, \widehat Z_-^{-1}(k)\bigr) ^{j-1}\, (V+B-\lambda )\, \Psi _{\zeta }\in {\mathcal H}^{n+2-j}_B
$$
and
$$
\| \Phi _{(j)}\| _{k,{\mathcal H}^{n+2-j}_B}\, \leqslant \, C_6(j)\, \bigl( 1+\| V+B-\lambda \| _{H^n_{\Lambda }}\bigr) ^j
\eqno(10)
$$
where $C_6(j)=C_6(\Lambda ,B,\widetilde K^*;j)>0$, $j=1,\dots ,n$. Furthermore, $\widetilde {\Phi }_{(n+1)}\in {\mathcal H}^2_B$, $\widehat P^{(0)}(k)\, \widetilde {\Phi }_{(n+1)}=0_B$ and
$$
(\widehat Z_+(k)\widehat Z_-(k)+V+B-\lambda )\, \Phi _{(n)}\, =\, -\, \zeta ^{-1}\, (\widehat Z_+(k)\widehat Z_-(k)+\zeta \widehat Z_-(k)+V+B-\lambda )\,\widetilde {\Phi }_{(n+1)}.
$$
Since $\| \Phi _{(n)}(\cdot |_{\widetilde K})\| _{C(\widetilde K)}\leqslant C_7(\Lambda ,B,\widetilde K^*)\| \Phi _{(n)}\| _{k,{\mathcal H}^2_B}$, it follows from (7) and (10) that
$$
\| (\widehat Z_+(k)\widehat Z_-(k)+V+B-\lambda )\, \Phi _{(n)}\| _{{\mathcal H}_B}\, \leqslant \, C_8(n)\, \bigl( 1+\| V+B-\lambda \| _{H^n_{\Lambda }}\bigr) ^{n+1}
$$
where $C_8(n)=C_8(\Lambda ,B,\widetilde K^*;n)>0$. On the other hand, Lemma 3 and Theorem 4 (if $|\zeta |\geqslant T(V+B-\lambda ;1/2)$) imply
$$
\| (\widehat Z_+(k)\widehat Z_-(k)+\zeta \widehat Z_-(k)+V+B-\lambda )\,\widetilde {\Phi }_{(n+1)}\| _{{\mathcal H}_B}\, \geqslant
$$ $$
\geqslant \, \frac 12\, \| (\widehat Z_+(k)\widehat Z_-(k)+\zeta \widehat Z_-(k))\,\widetilde {\Phi }_{(n+1)}\| _{{\mathcal H}_B}\, \geqslant \,
\frac {\sqrt B}{2\sqrt 2}\, |\zeta |\, \| \widetilde {\Phi }_{(n+1)}\| _{{\mathcal H}_B}.
$$
Therefore,
$$
\| \widetilde {\Phi }_{(n+1)}\| _{{\mathcal H}_B}\, \leqslant \, 2\sqrt 2\, B^{-1/2}\, C_8(n)\, \bigl( 1+\| V+B-\lambda \| _{H^n_{\Lambda }}\bigr) ^{n+1}.
\eqno(11)
$$
Let $\zeta =-Y_1+iY_2$ where $Y\in 2\pi \Lambda ^*$. The next estimate is a consequence of (11) and Lemma 1:
$$
|(e^{\. i(Y,x)}\Psi _{\zeta },(V+B-\lambda )\, \widetilde {\Phi }_{(n+1)})_{{\mathcal H}_B}|\, \leqslant \, \| (V+B-\lambda )\Psi _{\zeta }\| _{{\mathcal H}_B}\, \| \widetilde {\Phi }_{(n+1)}\| _{{\mathcal H}_B}\, 
\leqslant \eqno(12)
$$ $$
\leqslant \, C_1\sqrt Q\, \| V+B-\lambda \| _{L^2_{\Lambda }}\, \| \widetilde {\Phi }_{(n+1)}\| _{{\mathcal H}_B}\, \leqslant \, 2\sqrt {2Q}\, B^{-1/2}\, C_1C_8(n)\, \bigl( 1+\| V+B-\lambda \| _{H^n_{\Lambda }}\bigr) ^{n+2}.
$$
Now, let obtain (where $n\in {\mathbb N}$) the estimates for
$$
|(e^{\. i(Y,x)}\Psi _{\zeta },(V+B-\lambda )\, \Phi _{(j)})_{{\mathcal H}_B}|, \quad j=1,\dots ,n.
$$
Since $\Phi _{(j)}\in {\mathcal H}^{n+2-j}_B$ and $V+B-\lambda \in H^n_{\Lambda }({\mathbb R}^2;{\mathbb R})$, an appeal to Lemmas 4, 5 and estimate (10) (with some constants $C_9(j)=C_9(\Lambda ,B,\widetilde K^*;j)>0$) shows that
$$
\| (V+B-\lambda )\, \Phi _{(j)}\| _{k,{\mathcal H}^{n+1-j}_B}\, \leqslant \, C_9(j)\, \| V+B-\lambda \| _{H^n_{\Lambda }}\, \| 
\Phi _{(j)}\| _{k,{\mathcal H}^{n+2-j}_B}\, \leqslant
$$ $$
\leqslant \, C_6(j)\, C_9(j)\, \bigl( 1+\| V+B-\lambda \| _{H^n_{\Lambda }}\bigr) ^{j+1}.
$$
Hence (see (7)),
$$
\chi _j\, \doteq \, \widehat Z_+^{n+1-j}(k)\, (V+B-\lambda )\, \Phi _{(j)}\in {\mathcal H}_B ,
$$
and $\| \chi _j\| _{{\mathcal H}_B}=\| (V+B-\lambda )\, \Phi _{(j)}\| _{k,{\mathcal H}^{n+1-j}_B}$. On the other hand,
$$
\| \chi _j\| _{{\mathcal H}_B}\, \geqslant \, |(e^{\, i(Y,x)}\Psi _{\zeta },\chi _j)_{{\mathcal H}_B}|\, =\, |(\widehat Z_-^{n+1-j}(k)e^{\, i(Y,x)}\Psi _{\zeta }, (V+B-\lambda )\, \Phi _{(j)})_{{\mathcal H}_B}|\, =
$$ $$
=\, |Y|^{n+1-j}\, |(e^{\, i(Y,x)}\Psi _{\zeta },(V+B-\lambda )\, \Phi _{(j)})_{{\mathcal H}_B}|
$$
and therefore
$$
|(e^{\, i(Y,x)}\Psi _{\zeta },(V+B-\lambda )\, \Phi _{(j)})_{{\mathcal H}_B}|\, \leqslant \eqno(13)
$$ $$
\leqslant \, C_6(j)\, C_9(j)\, \bigl( 1+\| V+B-\lambda \| _{H^n_{\Lambda }}\bigr) ^{j+1}\, |\zeta |^{j-n-1},\quad j=1,\dots ,n.
$$
Since
$$
\bigl( (\widehat Z_+(k)+\zeta )\widehat Z_-(k)+V+B-\lambda \bigr) \, \Phi _{\zeta }\, =\, 0_B
$$
and
$$
(e^{\, i(Y,x)}\Psi _{\zeta },(\widehat Z_+(k)+\zeta )\widehat Z_-(k)\, \Phi _{\zeta })_{{\mathcal H}_B}\, =\, ((\widehat Z_-(k)+\overline {\zeta })e^{\, i(Y,x)}\Psi _{\zeta },\widehat Z_-(k)\, \Phi _{\zeta })_{{\mathcal H}_B}\, =\,
0,
$$
we have
$$
0\, =\, (e^{\, i(Y,x)}\Psi _{\zeta },(V+B-\lambda )\, \Phi _{\zeta})_{{\mathcal H}_B}\, =\,  
(e^{\, i(Y,x)}\Psi _{\zeta },(V+B-\lambda )\, \Psi _{\zeta })_{{\mathcal H}_B}\, +
$$ $$
+\, \sum\limits_{j\, =\, 1}^n\zeta ^{-j}\, (e^{\, i(Y,x)}\Psi _{\zeta },(V+B-\lambda )\, \Phi _{(j)})_{{\mathcal H}_B}\, +\, \zeta ^{-n-1}\, (e^{\, i(Y,x)}\Psi _{\zeta },(V+B-\lambda )\, \widetilde \Phi _{(n+1)})_{{\mathcal H}_B},
$$
and estimate (9) follows from (12) and (13). Theorem 5 is proved.
\vskip 0.2cm

Under the conditions of Theorem 5 (if $Y\in 2\pi \Lambda ^*$ and $|Y|\geqslant T(V+B-\lambda ;1/2)$) using the Lemma 2 we obtain
$$
C_5(n)\, (1+\| V+B-\lambda \| _{H^n_{\Lambda }})^{n+2}\geqslant |Y|^{n+1}\, |(e^{\, i(Y,x)}\Psi _{\zeta },(V+B-\lambda )\, \Psi _{\zeta })_{{\mathcal H}_B}|\, \geqslant
$$ $$
\geqslant |Y|^{n+1} \left( |V_Y|-\left| \, \sum\limits_{Y^{\prime }\, \in \, 2\pi \Lambda ^*\backslash \{ Y\} }V_{Y^{\prime }}\, (\Psi _{\zeta },e^{\, i(Y^{\prime }-Y,x)}\Psi _{\zeta })_{{\mathcal H}_B}\, \right| -|B-\lambda |\, \bigl|
(\Psi _{\zeta },e^{-i(Y,x)}\Psi _{\zeta })_{{\mathcal H}_B}\bigr| \right) \geqslant
$$ $$
\geqslant \, |Y|^{n+1} \left( |V_Y|\, -\, \sum\limits_{Y^{\prime }\, \in \, 2\pi \Lambda ^*\backslash \{ Y\} }|V_{Y^{\prime }}|\, e^{-|Y^{\prime }-Y|^2/(4B)}-|B-\lambda |\, e^{-|Y|^2/(4B)} \right) \geqslant
$$ $$
\geqslant \, |Y|^{n+1}\, {\mathcal C}_{B,V}(Y) - (B+|\lambda |)\, |Y|^{n+1}e^{-|Y|^2/(4B)}
$$
where $|Y|^{n+1}e^{-|Y|^2/(4B)}\leqslant (2B(n+1))^{(n+1)/2}\, e^{-(n+1)/2}$. Therefore the following theorem\footnote{In the formulation of the Theorem 6 there are no functions $\Psi _{\zeta }$ defined for $\zeta =-Y_1+iY_2$ and for the vector $k\in 2\pi \widetilde K^*$. Hence, we can assume that the numbers $C(n)$ and $T_0(n;V,\lambda )$ do not depend on the choice of the unit cells $\widetilde K$ and $\widetilde K^*$.} is a consequence of Theorem 5.
\vskip 0.2cm

{\bf Theorem 6}.
{\it For a given lattice $\Lambda \subset {\mathbb R}^2$ and for a homogeneous magnetic field with the flux $\eta =PQ^{-1}\in {\mathbb Q}$ there are numbers $C(n)=C(\Lambda ,B;n)>0$, $n\in {\mathbb Z}_+$ such that for every electric potential $V\in H^n_{\Lambda }({\mathbb R}^2;{\mathbb R})$ for which the operator $\widehat H_B+V$ has an eigenvalue $\lambda \in {\mathbb R}$, there exists a number $T_0(n;V,\lambda )=T_0(\Lambda ,B;n;V,\lambda )>0$ such that, for all vectors $Y\in 2\pi \Lambda ^*$ with $|Y|\geqslant T_0(n;V,\lambda )$, the estimate
$$
|Y|^{n+1}\, {\mathcal C}_{B,V}(Y)\, \leqslant \, C(n)\, \bigl( (v(K))^{(n+2)/2}\, (1+B+|\lambda |)^{n+2}+\| V\| _{H^n_{\Lambda }}^{n+2}\bigr) .
$$
holds.}
\vskip 0.2cm

Theorem 2 is a direct consequence of Theorem 6.
\vskip 0.2cm

{\bf Lemma 6}.
{\it For all $B>0$ and $Y\in 2\pi \Lambda ^*$, the function $W\mapsto {\mathcal C}_{B,W}(Y)\in {\mathbb R}$ is continuous in the space $(L^2_{\Lambda }({\mathbb R}^2;{\mathbb R}),\| \cdot \| _{L^2_{\Lambda }})$.} 
\vskip 0.2cm

Lemma 6 follows from the estimate
$$
|\, {\mathcal C}_{B,W}(Y)-{\mathcal C}_{B,W^{\prime }}(Y)\, |\leqslant \sum\limits_{Y^{\prime }\in \, 2\pi \Lambda ^*}|\, W_{Y^{\prime }}-W^{\prime }_{Y^{\prime }}|\, e^{-|Y^{\prime }-Y|^2/(4B)}\leqslant
$$ $$
\leqslant \, ((v(K))^{-1/2}\left( \, 
\sum\limits_{Y^{\prime }\in \, 2\pi \Lambda ^*}e^{-|Y^{\prime }|^2/(2B)}\right) ^{1/2}\| W-W^{\prime }\| _{L^2_{\Lambda }}, 
$$
which holds for all functions $W,W^{\prime }\in L^2_{\Lambda }({\mathbb R}^2;{\mathbb R})$.

\section{Proof of Theorem 1}

Let $S_j\in {\mathbb N}$, $j=1,2,3$. For a given lattice $\Lambda \subset {\mathbb R}^2$ and a homogeneous magnetic field with the flux $\eta =PQ^{-1}$ let's denote by ${\mathcal B}_n(S_1,S_2,S_3)={\mathcal B}_n(\Lambda ,B;S_1,S_2,S_3)$, $n\in {\mathbb Z}_+$, the set of potentials $V\in H^n_{\Lambda }({\mathbb R}^2;{\mathbb R})$ such that

(1)\, $\| V\| _{H^n_{\Lambda }}\leqslant S_1$,

(2)\, the operator $\widehat H_B+V$ has an eigenvalue $\lambda \in [-S_2,S_2]$,

(3)\, for all vectors $Y\in 2\pi \Lambda ^*$ with $|Y|\geqslant S_3$, the estimate
$$
|Y|^{n+1}\, {\mathcal C}_{B,V}(Y)\leqslant C^{\, \prime }(\Lambda ,B;n;S_1,S_2)\doteq C(\Lambda ,B;n)\, \bigl( (v(K))^{(n+2)/2}(1+B+S_2)^{n+2}+S_1^{n+2}\bigr)
$$
holds.
\vskip 0.2cm

{\bf Theorem 7}.
{\it For a given lattice $\Lambda \subset {\mathbb R}^2$ and a homogeneous magnetic field with the flux $\eta =PQ^{-1}$, potentials $V\in H^n_{\Lambda }({\mathbb R}^2;{\mathbb R})$, for which operator $\mathrm (1)$ has an eigenvalue $\lambda \in {\mathbb R}$, belong to the set
$$
{\mathcal B}_n(\Lambda ,B)\, \doteq \, \bigcup\limits_{S_1\, =\, 1}^{+\infty }\, \bigcup\limits_{S_2\, =\, 1}^{+\infty }\, \bigcup\limits_{S_3\, =\, 1}^{+\infty }{\mathcal B}_n(\Lambda ,B;S_1,S_2,S_3).
$$}
\vskip 0.2cm

Theorem 7 is a consequence of Theorem 6.
\vskip 0.2cm

{\bf Theorem 8}.
{\it For all $n\in {\mathbb Z}_+$ and all $S_j\in {\mathbb N}$, $j=1,2,3$, the set
$$
{\mathcal L}^n_{\Lambda }({\mathbb R}^2;{\mathbb R})\, \cap \, {\mathcal B}_n(\Lambda ,B;S_1,S_2,S_3)
$$
is nowhere-dense in the space $({\mathcal L}^n_{\Lambda }({\mathbb R}^2;{\mathbb R}).\| \cdot \| _{{\mathcal L}^n_{\Lambda }})$.}
\vskip 0.2cm

{\it Proof}.
It suffices to prove that for any function $W\in {\mathcal L}^n_{\Lambda }({\mathbb R}^2;{\mathbb R})\, \cap \, {\mathcal B}_n(S_1,S_2,S_3)$ and any $\varepsilon >0$ there exist a function $W^{\prime }\in {\mathcal L}^n_{\Lambda }({\mathbb R}^2;{\mathbb R})$ and a number $\widetilde {\varepsilon }>0$ such that $\| W-W^{\prime }\| _{{\mathcal L}^n_{\Lambda }}<\varepsilon $ and all functions $W^{\prime \prime }\in {\mathcal L}^n_{\Lambda }({\mathbb R}^2;{\mathbb R})$ with $\| W^{\prime }-W^{\prime \prime }\| _{{\mathcal L}^n_{\Lambda }}<\widetilde {\varepsilon }$ do not belong to the set ${\mathcal B}_n(S_1,S_2,S_3)$.

Let $W\in {\mathcal L}^n_{\Lambda }({\mathbb R}^2;{\mathbb R})\, \cap \, {\mathcal B}_n(S_1,S_2,S_3)$ and let $\varepsilon >0$. Then for all vectors $Y\in 2\pi \Lambda ^*$ with $|Y|\geqslant S_3$, the estimate $|Y|^{n+1}\, {\mathcal C}_{B,W}(Y)\leqslant C^{\, \prime }(\Lambda ,B;n;S_1,S_2)$ is fulfilled.

Let denote $a\doteq 4\pi \, (\ln 2)^{-3/4}\, {\mathrm {diam}}\, K^*$ where ${\mathrm {diam}}\, K^*$ is diameter of arbitrarily chosen unit cell $K^*$ of the lattice $\Lambda ^*$, and set
$$
R_m=am\, (\ln m)^{3/4},\quad r^{\prime }_m=\frac a2\, (\ln m)^{3/4},\quad r_m=\frac 12\, r^{\prime }_m,\quad m\in {\mathbb N}\backslash \{ 1\} .
$$
Then $R_{m+1}-R_m>r^{\prime }_{m+1}+r^{\prime }_m$ for all $m\in {\mathbb N}\backslash \{ 1\} $. Let denote
$$
{\mathcal P}_m(W)\, =\, \sum\limits_{Y^{\prime }\in \, 2\pi \Lambda ^*:\, R_m-r^{\prime }_m\leqslant \, |Y^{\prime }|\,\leqslant \, R_m+r^{\prime }_m}(1+|Y^{\prime }|)^{2m}\, |W_{Y^{\prime }}|^2
$$
and for vectors $x\in {\mathbb R}^2$ with $|x|=R_m$, let denote
$$
{\mathcal P}_m(W;x)\, =\, \sum\limits_{Y^{\prime }\in \, 2\pi \Lambda ^*:\, |Y^{\prime }-x|\, \leqslant \, r^{\prime }_m}(1+|Y^{\prime }|)^{2m}\, |W_{Y^{\prime }}|^2.
$$
Because 
$$
\sum\limits_{m\, \geqslant \, 2}
{\mathcal P}_m(W)\, \leqslant \, (v(K))^{-1}\, \| W\| ^2_{H^n_{\Lambda }}
$$
and the series $\sum\limits_{m\, \geqslant \, 2}(m\ln m)^{-1}$ is divergent, for any $\delta >0$ there is an infinite set ${\mathfrak M}(\delta )\subseteq {\mathbb N}\backslash \{ 1\} $ such that for all $m\in {\mathfrak M}(\delta )$ 
$$
{\mathcal P}_m(W)\, \leqslant \, \delta \, (v(K))^{-1}\, (m\ln m)^{-1}\, \| W\| ^2_{H^n_{\Lambda }} .
\eqno(14)
$$
In addition, for each $m\in {\mathfrak M}(\delta )$ there exists a vector $x\in {\mathbb R}^2$, $|x|=R_m$, such that
$$
{\mathcal P}_m(W;x)\, +\, {\mathcal P}_m(W;-x)\, \leqslant \, 2\pi ^{-1}\bigl( {\mathrm {arcsin}}\, r^{\prime }_m(R_m+r^{\prime }_m)^{-1}\bigr)\, {\mathcal P}_m(W).
\eqno(15)
$$
In view of inequality $r_m\geqslant \pi \, {\mathrm {diam}}\, K^*$, for this vector $x$ we can choose a vector $Y^{(m)}\in 2\pi \Lambda ^*$ with $|x-Y^{(m)}|\leqslant r_m$ (and $|x-Y^{\prime }|\leqslant r_m^{\prime }$ for all vectors $Y^{\prime }\in 2\pi \Lambda ^*$ with $|Y^{\prime }-Y^{(m)}|\leqslant r_m$). For all $m\in {\mathbb N}\backslash \{ 1\} $
$$
\frac 78\, am\, (\ln m)^{3/4}\, \leqslant \, |Y^{(m)}|\, \leqslant \, \frac 98\, am\, (\ln m)^{3/4}.
\eqno(16)
$$
Let denote
$$
W^{\, (Y^{(m)},r_m)}(x)\, = \sum\limits_{Y^{\prime }\in \, 2\pi \Lambda ^*:\, |Y^{\prime }-Y^{(m)}|\, \leqslant \, r_m}W_{Y^{\prime }}\, e^{\, i(Y^{\prime },x)},\quad x\in {\mathbb R}^2.
$$
By virtue of (14) and (15), the following lemma is valid.
\vskip 0.2cm

{\bf Lemma 7}.
{\it For every $\delta >0$ there is a number ${\delta }^{\, \prime }>0$ such that for all $m\in {\mathfrak M}({\delta }^{\, \prime })$ 
$$
\| W^{\, (Y^{(m)},r_m)}\| _{H^n_{\Lambda }}\, +\, \| W^{\, (-Y^{(m)},r_m)}\| _{H^n_{\Lambda }}\, \leqslant \, \delta \, m^{-1} (\ln m)^{-1/2}\, \| W\| _{H^n_{\Lambda }} .
$$} 
\vskip 0.2cm

We have $W^{\, (\pm Y^{(m)},r_m)}\in {\mathcal L}^n_{\Lambda }({\mathbb R}^2;{\mathbb C})$. Since for all $Y\in 2\pi \Lambda ^*$ and all $r\geqslant \pi \, {\mathrm {diam}}\, K^*$ the inequality
$$
{\mathcal N}\{ \{ Y^{\prime }\in 2\pi \Lambda ^*:|Y^{\prime }-Y|\leqslant r\} \}\leqslant (4\pi )^{-1} (v(K^*))^{-1} (r+\pi \, {\mathrm {diam}}\, K^*)^2\leqslant \pi ^{-1} v(K)\, r^2
$$
holds (where ${\mathcal N}\{ {\mathfrak G}\} $  is number of vectors in a finite set ${\mathfrak G}\subset 2\pi \Lambda ^*$), we obtain the estimate
$$
\| W^{\, (\pm Y^{(m)},r_m)}\| _{{\mathcal L}^n_{\Lambda }}\,
\leqslant \, {\mathcal C}^{\, \prime }_{{\mathcal L}^n_{\Lambda }}\sum\limits_{Y^{\prime }\in \, 2\pi \Lambda ^*:\, |Y^{\prime }\mp \, Y^{(m)}|\, \leqslant \, r_m}
(1+|Y^{\prime }|)^{n+\theta }\, |W_{Y^{\prime }}|\, \leqslant
\eqno(17)
$$ $$
\leqslant \, {\mathcal C}^{\, \prime }_{{\mathcal L}^n_{\Lambda }}\, (1+|Y^{(m)}|+r_m)^{\theta }({\mathcal N}\{ \{ Y^{\prime }\in 2\pi \Lambda ^*:|Y^{\prime }\mp \, Y^{(m)}|\leqslant r_m\} \} )^{1/2}\, \times
$$ $$
\times \, 
(v(K))^{-1/2}\, \| W^{\, (\pm Y^{(m)},r_m)}\| _{H^n_{\Lambda }}\, \leqslant
$$ $$
\leqslant \, \frac {a}{4\sqrt {\pi }}\, \left( \frac 98\right) 
^{\theta }{\mathcal C}^{\, \prime }_{{\mathcal L}^n_{\Lambda }}\, (1+|Y^{(m)}|)^{\theta }\, (\ln m)^{3/4}\, \| W^{\, (\pm Y^{(m)},r_m)}\| _{H^n_{\Lambda }},\quad m\in {\mathbb N}\backslash \{ 1\} .
$$
\vskip 0.2cm

{\bf Lemma 8}.
{\it For every $\delta >0$ there is a number ${\delta }^{\, \prime \prime }>0$ such that for all $m\in {\mathfrak M}({\delta }^{\, \prime \prime })$ 
$$
\| W^{\, (Y^{(m)},r_m)}\| _{{\mathcal L}^n_{\Lambda }}\, +\, \| W^{\, (-Y^{(m)},r_m)}\| _{{\mathcal L}^n_{\Lambda }}\, \leqslant \, \delta \, m^{\theta -1} (\ln m)^{(1+3\theta )/4}\, \| W\| _{H^n_{\Lambda }} .
$$} 
\vskip 0.2cm

Lemma 8 follows from Lemma 7 and estimate (17).

For all $m\in {\mathbb N}\backslash \{ 1\} $, we define functions 
$$
W^{(m)}(x)\doteq W(x)-W^{\, (Y^{(m)},r_m)}(x)-W^{\, (-Y^{(m)},r_m)}(x)\, +
$$ $$
+\, m^{-n-(1+\theta )/2}\, \bigl( e^{\, i(Y^{(m)},x)}+e^{- i(Y^{(m)},x)}
\bigr) ,\quad x\in {\mathbb R}^2,
$$
which belong to the space ${\mathcal L}^n_{\Lambda }({\mathbb R}^2;{\mathbb R})$. Let the number ${\delta }^{\, \prime \prime }$ be determined for the number $\delta =1$ in Lemma 8. By Lemma 8 and estimates (16), for all $m\in {\mathfrak M}({\delta }^{\, \prime \prime })$ 
$$
\| W-W^{(m)}\| _{{\mathcal L}^n_{\Lambda }}\, \leqslant \, \| W^{\, (Y^{(m)},r_m)}\| _{{\mathcal L}^n_{\Lambda }}\, +\, \| W^{\, (-Y^{(m)},r_m)}\| _{{\mathcal L}^n_{\Lambda }}\, +
\eqno(18)
$$ $$
+\, 2\, {\mathcal C}^{\, \prime }_{{\mathcal L}^n_{\Lambda }}\, 
m^{-n-(1+\theta )/2}\, (1+|Y^{(m)}|)^{n+\theta }\, \leqslant
$$ $$
\leqslant \, m^{\theta -1} (\ln m)^{(1+3\theta )/4}\, \| W\| _{H^n_{\Lambda }}\, +\, 2\, {\mathcal C}^{\, \prime }_{{\mathcal L}^n_{\Lambda }}\, 
m^{-n-(1+\theta )/2}\, \left( 1+\frac98\, am\, (\ln m)^{3/4}\right ) ^{n+\theta }.
$$
For $m\in {\mathbb N}\backslash \{ 1\} $,
$$
{\mathcal C}_{B,W^{(m)}}(Y^{(m)})\geqslant m^{-n-(1+\theta )/2}\, \bigl( 1-e^{-|Y^{(m)}|^2/B}\bigr) \, - \sum\limits_{Y^{\prime }\in \, 2\pi \Lambda ^*:\, |Y^{\prime }- \, Y^{(m)}|\, > \, r_m}
|W_{Y^{\prime }}|\, e^{-|Y^{\prime }-Y^{(m)}|^2/(4B)}
$$
and since 
$|W_{Y^{\prime }}|\, \leqslant \, (v(K))^{-1/2}\, \| W\| _{L^2_{\Lambda }}$,  $Y^{\prime }\in 2\pi \Lambda ^*$,
and
$$
\sum\limits_{Y^{\prime }\in \, 2\pi \Lambda ^*:\, |Y^{\prime }|\, > \, r_m}e^{-|Y^{\prime }|^2/(4B)}\, \leqslant \, C^{\, \prime \prime }e^{-r_m^2/(8B)}, \quad \quad m\in {\mathbb N}\backslash \{ 1\} ,
$$
where $C^{\, \prime \prime }=C^{\, \prime \prime }(\Lambda ,B)>0$, we derive (also see (16))
$$
|Y^{(m)}|^{n+1}{\mathcal C}_{B,W^{(m)}}(Y^{(m)})\, \geqslant \eqno(19)
$$ $$
\geqslant \, \left( \frac 78\right) ^{n+1}a^{n+1}(\ln m)^{3(n+1)/4}\Bigl( m^{(1-\theta )/2}\bigl( 1-e^{-(49/64)a^2B^{-1}m^2(\ln m)^{3/2}}\bigr) \, -
$$ $$
-\, C^{\, \prime \prime }(v(K))^{-1/2}\, \| W\| _{L^2_{\Lambda }}\, m^{n+1}e^{-(1/128)a^2B^{-1}(\ln m)^{3/2}}\Bigr) .
$$
From (16), (18), and (19) it follows that for a sufficiently large number $m\in {\mathfrak M}({\delta }^{\, \prime \prime })$ the inequalities $\| W-W^{(m)}\| _{{\mathcal L}^n_{\Lambda }}<\varepsilon $ and $|Y^{(m)}|\geqslant S_3$, $|Y^{(m)}|^{n+1}{\mathcal C}_{B,W^{(m)}}(Y^{(m)})>C^{\, \prime }(\Lambda ,B;n;S_1,S_2)$ hold. Therefore, $W^{(m)}\notin {\mathcal B}_n(S_1,S_2,S_3)$. According to Lemma 6, there exists a number $\widetilde {\varepsilon }>0$ such that for all functions $W^{\prime \prime }\in {\mathcal L}^n_{\Lambda }({\mathbb R}^2;{\mathbb R})$ with $\| W^{\prime \prime }-W^{(m)}\| _{{\mathcal L}^n_{\Lambda }}<\widetilde {\varepsilon }$, the inequality $|Y^{(m)}|^{n+1}{\mathcal C}_{B,W^{\prime \prime }}(Y^{(m)})>C^{\, \prime }(\Lambda ,B;n;S_1,S_2)$ also fulffiled and consequently $W^{\prime \prime }\notin {\mathcal B}_n(S_1,S_2,S_3)$. Hence, the set ${\mathcal L}^n_{\Lambda }({\mathbb R}^2;{\mathbb R})\cap {\mathcal B}_n(S_1,S_2,S_3)$ is nowhere-dense in the space ${\mathcal L}^n_{\Lambda }({\mathbb R}^2;{\mathbb R})$. Theorem 8 is proved. 
\vskip 0.2cm

{\it Proof} of Theorem 1. It follows from Theorem 8 that for a homogeneous magnetic field with the flux $\eta =PQ^{-1}\in {\mathbb Q}$, the set ${\mathcal L}^n_{\Lambda }({\mathbb R}^2;{\mathbb R})\cap {\mathcal B}_n(\Lambda ,B)$ is of first Baire category in the space $({\mathcal L}^n_{\Lambda }({\mathbb R}^2;{\mathbb R}), \| \cdot \| _{{\mathcal L}^n_{\Lambda }})$. Because ${\mathbb Q}$ is a countable set, the union of the sets ${\mathcal L}^n_{\Lambda }({\mathbb R}^2;{\mathbb R})\cap {\mathcal B}_n(\Lambda ,B)$
over all coprime numbers $P,Q\in {\mathbb N}$ is also a set ${\mathfrak B}$ of first Baire category, and therefore its complement ${\mathcal L}^n_{\Lambda }({\mathbb R}^2;{\mathbb R})\backslash {\mathfrak B}$ containes a dense $G_{\delta}$-set. If operator (1), for a potential $V\in {\mathcal L}^n_{\Lambda }({\mathbb R}^2;{\mathbb R})\subseteq H^n_{\Lambda }({\mathbb R}^2;{\mathbb R})$ and for a homogeneous magnetic field with the flux $0<\eta \in {\mathbb Q}$, has an eigenvalue $\lambda \in {\mathbb R}$, then Theorem 7 implies that $V\in {\mathfrak B}$. Hence, for all potentials $V\in {\mathcal L}^n_{\Lambda }({\mathbb R}^2;{\mathbb R})\backslash {\mathfrak B}$ and for all homogeneous magnetic fields with the flux $0<\eta \in {\mathbb Q}$, the spectrum of the operator $\widehat H_B+V$ has no eigenvalues and, consequently, the spectrum is absolutely continuous. Theorem 1 is proved.

\end{document}